\documentclass[12pt]{article}
\usepackage{jheppub}
\usepackage{epsfig}
\usepackage{amsfonts}
\usepackage{amssymb}
\usepackage{enumitem}
\usepackage{amsthm}
\usepackage{subfig}
\usepackage{bm}
\usepackage{hyperref}
\usepackage{mathrsfs}
\usepackage{bbm}
\usepackage{cancel}
\usepackage[dvipsnames]{xcolor}
\usepackage{accents}
\usepackage{relsize}
\usepackage{float, slashed, graphicx, amssymb, amsmath}
\usepackage{shuffle}
\usepackage{tikz}
\usetikzlibrary{hobby,calc,shapes, positioning, backgrounds,trees, decorations, decorations.markings, decorations.pathmorphing, arrows, shapes.geometric, snakes}
\usepackage{tikz-feynman}
\tikzfeynmanset{compat=1.1.0}
\tikzfeynmanset{
graviton/.style={circle, draw=green!60, fill=green!5, very thick, minimum size=7mm}
}
\tikzfeynmanset{
codot/.style={/tikz/shape=circle,
/tikz/fill=red,/tikz/minimum size=0.1cm,/tikz/inner sep=1.8pt}
}
\tikzfeynmanset{sb/.style=
{/tikz/shape=circle,
/tikz/fill=black,
 /tikz/minimum size=0.3cm,/tikz/inner sep=1.pt
 } }
\tikzfeynmanset{myblob/.style=
{/tikz/shape=ellipse,
/tikz/fill=red,
 /tikz/minimum width=0.5cm,
 } }
 \tikzfeynmanset{myblobFF/.style=
{/tikz/shape=circle,
/tikz/fill=black,
 /tikz/minimum width=0.25cm,
 } }
 \tikzfeynmanset{myblob2/.style=
{/tikz/shape=rectangle,
/tikz/fill=red,
 /tikz/minimum width=0.1cm,/tikz/inner sep=1.8pt
 } }
  \tikzfeynmanset{myvertex/.style=
{/tikz/shape=circle,
/tikz/fill=black,
 /tikz/minimum width=0.05cm,
 } }
 \tikzfeynmanset{GR/.style={/tikz/shape=ellipse,/tikz/fill={rgb:black,1;white,2},/tikz/minimum size=0.3cm,} }
\tikzset{box/.pic={\filldraw[fill=black]  (0,0) circle (2.5pt);
				   \filldraw [fill=black] (0.5,0) circle (2.5pt);
			       \draw [line width=5pt] (0,0) -- (0.5,0);}}

\tikzset{wiggle/.style={decorate, decoration=snake}}

 \tikzfeynmanset{HV/.style=
{/tikz/shape=circle,
/tikz/fill={rgb:black,1;white,2},
 /tikz/minimum size=0.01cm,/tikz/inner sep=1.8pt
 } }

\usepackage[T1]{fontenc} 

\newcommand{\tr}{\text{tr}}

\def\sc#1{\overline{#1}}
\makeatletter
\newcommand \UPlus {\mathop {\operator@font \uplus }\limits }
\makeatother
\makeatletter
\newcommand \Bigcup {\mathop {\operator@font \bigcup }\limits }
\makeatother
  \def\LabelNote#1{}
 \def\Label#1{\label{#1}%
  \smash{\hbox to\phipt{\raise1ex\hbox{\tiny[#1]}\hss}}}

\def\veps{\varepsilon}

\newcommand{\sinch}[1]{{\sinh(#1)\over #1} }

\def\nn{\nonumber}

\newcommand{\red}{\color{red}}

\newcommand{\black}{\color{black}}

\newcommand{\pink}{\color{Magenta}}

\def\Eq#1{Eq.~({\pink\ref{#1}})}

\def\spa#1.#2{\left\langle#1\,#2\right\rangle}
\def\spb#1.#2{\left[#1\,#2\right]}

\def\be{\begin{equation}}
\def\ee{\end{equation}}
\def\bea{\begin{eqnarray}}
\def\eea{\end{eqnarray}}  
\allowdisplaybreaks

\usepackage[normalem]{ulem}
\newcommand{\npre}{\mathcal{N}}

\newcommand{\floor}[1]{\lfloor #1 \rfloor}
  
  \title{{{Classical Spin Gravitational Compton Scattering}}}
  \author{ N.~E.~J.~Bjerrum-Bohr, Gang~Chen, Marcos~Skowronek}
\affiliation{Niels Bohr International Academy,
Niels Bohr Institute,\\ University of Copenhagen,
Blegdamsvej 17, \\DK-2100 Copenhagen \O, Denmark}
\emailAdd{ bjbohr@nbi.dk; gang.chen@nbi.ku.dk; marcos.santos@nbi.ku.dk}

\begin{document} 

\abstract{ 
We introduce a novel bootstrap method for heavy-mass effective field theory classical Compton scattering amplitudes involving two massless particles and two arbitrary-spin infinite-mass limit particles. Using a suitable ansatz, we deduce new and explicit classical spin results for gluon four and five-point infinite mass processes that exhibit a certain exponentiated three-point tree-level factorizations and feature no spurious poles. We discuss the generalization of our bootstrap to higher multiplicities and summarize future potential applications.}  

\keywords{Scattering amplitudes, color-kinematics duality and the double copy, general relativity from amplitudes}

\maketitle

\section{Introduction}
Observations of gravitational waves from black hole mergers have fast evolved into an exciting avenue for testing Einstein's classical theory. To facilitate this, a modern and efficient computational program inspired by various practical quantum field theory implementations of general relativity, see, for instance,~\cite{Iwasaki:1971vb,Donoghue:1994dn,Bjerrum-Bohr:2002gqz,Holstein:2004dn,Holstein:2008sx,Neill:2013wsa,Bjerrum-Bohr:2013bxa}, was developed in \cite{Bjerrum-Bohr:2018xdl,Cheung:2018wkq} prompted by the state of current theoretical analysis and precision requirements of observations \cite{Damour:2017zjx}. It has made quantum scattering amplitudes for superheavy black holes -- taken as point particles --  a powerful resource for delivering much-needed analytic precision in fully relativistic post-Minkowskian frameworks and promises new computational efficiency, particularly in analytical regions of the near-merger regimes.  \\[5pt]
Several post-Minkowskian spinless calculations have already made new landmarks by extending the expansion in Newton's constant $G_N$~\cite{Kosower:2018adc,Cristofoli:2019neg,Bern:2019nnu,Antonelli:2019ytb,Bern:2019crd,Parra-Martinez:2020dzs,DiVecchia:2020ymx,Damour:2020tta,Kalin:2020fhe,DiVecchia:2021ndb,DiVecchia:2021bdo,Herrmann:2021tct,Bjerrum-Bohr:2021vuf,Bjerrum-Bohr:2021din,Damgaard:2021ipf,Brandhuber:2021eyq,Bjerrum-Bohr:2021wwt,Bjerrum-Bohr:2022ows,Dlapa:2021npj,Bern:2021yeh,Bern:2022jvn} with~\cite{Bern:2021dqo} pioneering fourth post-Minkowskian order. Another research attraction is the inclusion of spin effects which can lead to significant quantitative augmentations of observed signals for events. For spinning black holes, improvement till now has been gradual because of the more complicated and specialized framework, but one finds several vital advancements using spinor helicity formalism in ~\cite{Guevara:2017csg,Arkani-Hamed:2017jhn,Guevara:2018wpp,Chung:2018kqs,Guevara:2019fsj,Arkani-Hamed:2019ymq,Aoude:2020onz,Chung:2020rrz,Guevara:2020xjx,Chen:2021kxt,Kosmopoulos:2021zoq,Chiodaroli:2021eug,Bautista:2021wfy,Cangemi:2022bew,Ochirov_2022,Comberiati:2022ldk}, using covariant and world-line field theory in~\cite{Damgaard:2019lfh,Bern:2020buy,Mogull:2020sak,Jakobsen:2021zvh,Dlapa:2021vgp,Dlapa:2022lmu} and with various phenomenological applications in \cite{Vines:2017hyw,Vines:2018gqi,Maybee:2019jus,Haddad:2021znf,Jakobsen:2021smu,Liu:2021zxr,Chen:2022clh,Jakobsen:2022fcj,Menezes:2022tcs,FebresCordero:2022jts,Alessio:2022kwv,Bern:2022kto,Aoude:2022thd,Aoude:2022trd}.  \\[5pt]
For addressing classical spins, it is customary to operate in the context of a Lagrangian formalism that involves higher spin particles systematized in expansions with increasing multi-pole interactions as pioneered in \cite{Singh:1974qz}. Unfortunately, the formalism behind this is relatively elaborate, and derived amplitudes usually contain intricate redundancies. In this presentation, we take an alternative route and bootstrap covariant arbitrary-spin amplitudes from factorization limits.  \\[5pt]
Our main focus here will be classical spin Compton scattering amplitudes, where an infinite mass spinning particle interacts with gluons. As we explain, we can extend such amplitudes into results for gravitational Compton scattering amplitudes utilizing the double-copy~\cite{Bern:2008qj,Bern:2010ue,Bern:2019prr}. The rationale for mainly considering gravitational Compton scattering amplitudes in the infinite mass limit is that such amplitudes (in the limit where the mass is infinite) are convenient building blocks in generalized unitarity approaches \cite{Damgaard:2021ipf,Brandhuber:2021eyq,Bjerrum-Bohr:2021wwt,Travaglini:2022uwo,Bjerrum-Bohr:2022blt, Bjerrum-Bohr:2022ows} that identify the gravitational classical radial action computing from an integrand determined by generalized unitarity and constrained by delta functions \cite{Brandhuber:2021eyq,Bjerrum-Bohr:2021vuf,Bjerrum-Bohr:2021din}. \\[5pt]
In the bootstrap method, we use minimal physical input constraints such as the special exponentiated minimally coupled Kerr black hole three-point amplitude {\red \cite{Arkani-Hamed:2017jhn,Guevara:2018wpp,Chung:2018kqs,Chen:2021kxt}}, and we will introduce a convenient ansatz for the amplitude that avoids spurious poles and is fixed uniquely (up to polynomial contact terms) from imposing factorization identities. We find support for our classical amplitudes by comparing them to results derived at four-point linear and quadratic spin order~\cite{Bern:2020buy,Bautista:2022wjf}. Beyond quadratic spin order, we stress that our formalism requires additional contact deformations to describe Kerr black hole scattering. We will return to this point in the conclusion.\\[5pt]
We arrange this paper as follows: in section \ref{section 2}, we will quickly review infinite mass Compton gluon and graviton tree amplitudes to establish a convenient and efficient framework for computation. Then we will introduce our new bootstrap method and demonstrate the calculation of four and five-point classical amplitudes in section \ref{section 3}. An essential inspiration in the bootstrap method is the formalism \cite{Brandhuber:2021bsf}. Finally, we will outline a generalization of the proposed computational technology in section \ref{section 4} and discuss possible applications and questions for the future in section \ref{section 5}.\\[5pt]

\section{Classical Compton amplitudes}\label{section 2}
The tree amplitudes considered in this paper involve an infinity mass and spin particle with finite classical ring radius $a=s/m$ ($s\rightarrow \infty$, $m\rightarrow \infty$), which interacts with an arbitrary $n-2$ number of gluons in the context of a heavy-mass effective field theory (HEFT) \cite{Georgi:1990um,Luke:1992cs,Neubert:1993mb,Manohar:2000dt,Damgaard:2019lfh,Brandhuber:2021bsf,Brandhuber:2021eyq,Brandhuber:2021kpo}. We depict the processes considered in the following way:
    \begin{align}
\label{compton}
& \begin{tikzpicture}[baseline={([yshift=-0.8ex]current bounding box.center)}]\tikzstyle{every node}=[font=\small]	
\begin{feynman}
    	 \vertex (a) {\( \bar p_n=mv\)};
    	 \vertex [right=2.1cm of a] (f2)[GR]{$~~~$$~~~$};
    	 \vertex [right=2.5cm of f2] (c){$\bar p_{n-1}=mv-q$};
    	 \vertex [above=1.3cm of f2] (gm){};
    	 \vertex [left=0.8cm of gm] (g2){$~~~~~~~~~ p_{1}~~~\cdots$};
    	  \vertex [right=0.8cm of gm] (g20){$ p_{n-2}$};
    	  \diagram* {
(a) -- [fermion,thick] (f2) --  [fermion,thick] (c),
    	  (g2)--[gluon,thick,rmomentum'](f2),(g20)--[gluon,thick,rmomentum](f2)
    	  };
    \end{feynman}  
    \end{tikzpicture}
\end{align}
Here $p_1,\ldots,p_{n-2}$ denotes momenta of gluons while $\bar p_{n-1}$ and $\bar p_n$ are momenta for the heavy spinning particle.  In the heavy mass limit the velocity $v$ is fixed by the on-shell condition  $v\cdot v=1$ and $v\cdot q=0$ which follows from $\bar p_{n}^2 = \bar p_{n-1}^2 = (m v-q)^2  =  m^2$.  We define $q\equiv ( p_{1}+\ldots+ p_{n-2})$.\\[5pt]

\subsection{Three-point amplitude}
We start with the infinite mass three-point Compton amplitude, which we define as follows (using shorthand notation for momenta)
\begin{align}\label{eq:Amp3}
\lim_{m\to \infty}A_a(1,\bar 2,\bar 3) \equiv A_a(1,v)\,.
\end{align}
It can conveniently be written in terms of the three-point color-kinematic numerator ${\npre_a(1,v)}$ 
\begin{align}\label{eq:npr}
&A_a(1,v)={\npre_a(1,v)}\,.
\end{align}
 We will in the following focus on the minimally coupled three-point tree amplitude that takes a closed exponentiated form for arbitrary spin and double copies to a Kerr graviton amplitude as shown in \cite{Arkani-Hamed:2017jhn,Guevara:2018wpp,Chung:2018kqs,Chen:2021kxt}
\begin{equation}\label{eq:3pt_pre}
 \npre_a(1,v) =m\, (v\cdot \veps_1)\, \exp\left(\, i\,\,{p_1\cdot S\cdot \veps_1\over m\, v\cdot \veps_1}\right).
\end{equation}
We can rewrite the exponential by separating the even and odd powers in the spin variable and using the identity $\epsilon^{\mu\nu\rho\sigma}\epsilon_{\alpha\beta\gamma\delta} = -\delta^{\mu\nu\rho\sigma}_{\alpha\beta\gamma\delta}$ 
\begin{equation}
    \begin{gathered}
       m\, v^\mu \exp\left(\, i\,\,{p_1\cdot S\cdot \veps_1\over m \,v\cdot \veps_1}\right) \doteq m \,\cosh(p_1\cdot a) v^{\mu} - \, i \,{} \,\Big({\sinh(p_1\cdot a)\over p_1\cdot a}\Big) (p_1\cdot S)^{\mu}\,,
\label{defw}    \end{gathered}
\end{equation}
where the $\doteq$ indicates that the equality is satisfied when contracted with $\varepsilon_1$. In the above expressions $\veps_i$ denotes gluon polarizations and $S^{\mu\nu}\equiv-m \epsilon^{\mu\nu\rho\sigma}v_\rho a_\sigma$. Since the singularity in the second term of Eq. (\ref{defw}) is removable, we can define an analytic extension that is holomorphic everywhere in the complex plane by considering the function
\begin{equation}\label{G1def}
\displaystyle G_1(x_1)= {\sinh(x_1)\over x_1}\ \ \forall \ \ x_1 \neq 0 \quad {\rm with}\ \ G_1(0)= \lim_{x_1\to 0} {\sinh(x_1)\over x_1} = 1.
\end{equation} 
Thus we can define  
\begin{equation}
    \begin{gathered}
        w_1^{\mu}\equiv m\,\cosh(p_1\cdot a) v^{\mu} - \, i \,{} \,\Big(G_1(p_1\cdot a)\Big) (p_1\cdot S)^{\mu}\,,
    \end{gathered}
\end{equation}
which defines a three-point numerator which has the advantage of manifestly featuring no pole terms in $a\cdot p_1$
\begin{equation}\label{eq:3pt}
 \npre_a(1,v) = w_1\cdot \veps_1.
\end{equation} 
We note that gauge invariance is manifest in \Eq{eq:3pt} from the anti-symmetric property of the spin tensor and the three point on-shell condition $v\cdot p_1=0$, which imply that $w_1\cdot p_1 =0$.
We conclude this section by noting that three-point gravity amplitudes can be computed from the double copy using a product of one scalar gluon numerator, $\npre_0(1,v)$, and one arbitrary spin numerator factor, $\npre_a(1,v)$ \cite{Johansson:2019dnu},
\begin{align}
M_a(1,v)=\npre_0(1,v)\,\,\npre_a(1,v)\,= m\,(v\cdot \veps_1)\,\, (w_1\cdot \veps_1)\,.
\end{align}
\subsection{Classical gravity amplitude decomposition in the heavy mass limit}\label{AmpDecomp}
Before we discuss the four-point gravity amplitude, we will review some material related to convenient amplitude decompositions in the heavy mass limit. We will here refer to the discussion in \cite{Brandhuber:2021eyq} (see also \cite{Bjerrum-Bohr:2021wwt}), and start by noting that it is convenient to decompose gravity amplitudes in the following way when considering classical black hole scattering perturbatively:
\begin{align}\label{decomp4}
\begin{tikzpicture}[baseline={([yshift=-0.8ex]current bounding box.center)}]\tikzstyle{every node}=[font=\small]	
\begin{feynman}
    	 \vertex (a) {\( \bar 4\)};
    	 \vertex [right=1.5cm of a] (f2)[GR]{$~~~~$};
    	 \vertex [right=1.5cm of f2] (c){$\bar 3$};
    	 \vertex [above=1.3cm of f2] (gm){};
    	 \vertex [left=0.8cm of gm] (g2){$1$};
    	  \vertex [right=0.8cm of gm] (g20){$2$};
    	  \diagram* {
(a) -- [fermion,thick] (f2) --  [fermion,thick] (c),
    	  (g2)--[photon,ultra thick](f2),(g20)--[photon,ultra thick](f2)
    	  };
    \end{feynman}  
    \end{tikzpicture}\xrightarrow[]{m\rightarrow \infty}
    \begin{tikzpicture}[baseline={([yshift=-0.4ex]current bounding box.center)}]\tikzstyle{every node}=[font=\small]
		\begin{feynman}
			\vertex (p1) {\(\sc 4\)};
			\vertex [right=0.8cm of p1] (b1) [dot]{};
			\vertex [right=0.5cm of b1] (b2) []{};
			\vertex [right=0.5cm of b2] (b3) [dot]{};
   \vertex [above=1.2cm of b1] (u1) []{$1$}; \vertex [above=1.2cm of b3] (u3) []{$2$};
			\vertex [right=0.8cm of b3](p4){$\sc 3$};
			\vertex [right=0.5cm of b1] (cut1);
			\vertex [above=0.2cm of cut1] (cut1u);
			\vertex [below=0.2cm of cut1] (cut1b);
			\vertex [right=0.25cm of b2] (cut2);
			\vertex [above=0.2cm of cut2] (cut2u);
			\vertex [below=0.2cm of cut2] (cut2b);
			\diagram* {
			(b1)--[photon,ultra thick](u1), (b3)-- [photon,ultra thick] (u3), (p1) -- [thick] (b1)-- [thick] (b3)-- [thick] (p4), (cut1u)--[ red,thick] (cut1b),
			};
		\end{feynman}
  \end{tikzpicture}+\begin{tikzpicture}[baseline={([yshift=-0.8ex]current bounding box.center)}]\tikzstyle{every node}=[font=\small]	
\begin{feynman}
    	 \vertex (a) {\( \bar 4\)};
    	 \vertex [right=1.2cm of a] (f2)[HV]{H};
    	 \vertex [right=1.2cm of f2] (c){$\bar 3$};
    	 \vertex [above=1.3cm of f2] (gm){};
    	 \vertex [left=0.8cm of gm] (g2){$1$};
    	  \vertex [right=0.8cm of gm] (g20){$2$};
    	  \diagram* {
(a) -- [fermion,thick] (f2) --  [fermion,thick] (c),
    	  (g2)--[photon,ultra thick](f2),(g20)--[photon,ultra thick](f2)
    	  };
    \end{feynman}  
    \end{tikzpicture},
\end{align}
The first term in this diagrammatical expression corresponds to a velocity cut imposed by a $\delta$-function constraint. This contribution is of order $\mathcal{O}(m^3)$\footnote{\footnotesize{Equivalently, if we count in powers of $\hbar$, of order $\mathcal{O}(\hbar^{-1})$}}. The second term, which we will refer to as the classical tree amplitude, is of order $\mathcal{O}(m^2)$ ($\mathcal{O}(\hbar^0)$). Since the first term is an iteration in the infinite mass limit, the following section focuses solely on computing the second term of (\ref{decomp4}). In Appendix A, we will discuss calculating the first contribution directly when considering a finite heavy mass.\\[5pt]
One can follow the same reasoning at five-point, and focus only on the last term of the following expansion:
\begin{align}\label{decomp}
\begin{tikzpicture}[baseline={([yshift=-0.8ex]current bounding box.center)}]\tikzstyle{every node}=[font=\small]	
\begin{feynman}
    	 \vertex (a) {\( \bar 5\)};
    	 \vertex [right=1.5cm of a] (f2)[GR]{$~~~~$};
    	 \vertex [right=1.5cm of f2] (c){$\bar 4$};
    	 \vertex [above=1.3cm of f2] (gm){$2$};
    	 \vertex [left=0.8cm of gm] (g2){$1$};
    	  \vertex [right=0.8cm of gm] (g20){$3$};
    	  \diagram* {
(a) -- [fermion,thick] (f2) --  [fermion,thick] (c),
    	  (g2)--[photon,ultra thick](f2), (gm)--[photon,ultra thick](f2),(g20)--[photon,ultra thick](f2)
    	  };
    \end{feynman}  
    \end{tikzpicture}&\xrightarrow[]{m\rightarrow \infty}\begin{tikzpicture}[baseline={([yshift=-0.4ex]current bounding box.center)}]\tikzstyle{every node}=[font=\small]
		\begin{feynman}
			\vertex (p1) {\(\sc 5\)};
			\vertex [right=0.8cm of p1] (b1) [dot]{};
			\vertex [right=0.5cm of b1] (b2) []{};
			\vertex [right=0.5cm of b2] (b3) [dot]{};
   \vertex [right=0.5cm of b3] (b4) []{};
   \vertex [right=0.5cm of b4] (b5) [dot]{};
   \vertex [above=1.2cm of b1] (u1) []{$1$}; \vertex [above=1.2cm of b3] (u3) []{$2$};
    \vertex [above=1.2cm of b5] (u5) []{$3$};
			\vertex [right=0.8cm of b5](p4){$\sc 4$};
			\vertex [right=0.5cm of b1] (cut1);
			\vertex [above=0.2cm of cut1] (cut1u);
			\vertex [below=0.2cm of cut1] (cut1b);
			\vertex [right=0.5cm of b3] (cut2);
			\vertex [above=0.2cm of cut2] (cut2u);
			\vertex [below=0.2cm of cut2] (cut2b);
   \vertex [right=0.5cm of b3] (cut3);
			\vertex [above=0.2cm of cut3] (cut3u);
			\vertex [below=0.2cm of cut3] (cut3b);
			\diagram* {
			(b1)--[photon,ultra thick](u1), (b3)-- [photon,ultra thick] (u3),(b5)-- [photon,ultra thick] (u5), (p1) -- [thick] (b1)-- [thick] (b3)-- [thick] (p4), (cut1u)--[ red,thick] (cut1b),(cut3u)--[ red,thick] (cut3b),
			};
		\end{feynman}
  \end{tikzpicture}+
    \begin{tikzpicture}[baseline={([yshift=-0.4ex]current bounding box.center)}]\tikzstyle{every node}=[font=\small]
		\begin{feynman}
			\vertex (p1) {\(\sc 5\)};
			\vertex [right=0.8cm of p1] (b1) [dot]{};
			\vertex [right=0.5cm of b1] (b2) []{};
			\vertex [right=0.5cm of b2] (b3) [HV]{H};
   \vertex [above=1.2cm of b1] (u1) []{$1$}; \vertex [above=1.2cm of b3] (u3) []{};\vertex [left=0.5cm of u3] (u3a) []{$2$};\vertex [right=0.5cm of u3] (u3b) []{$3$};
			\vertex [right=0.8cm of b3](p4){$\sc 4$};
			\vertex [right=0.5cm of b1] (cut1);
			\vertex [above=0.2cm of cut1] (cut1u);
			\vertex [below=0.2cm of cut1] (cut1b);
			\vertex [right=0.25cm of b2] (cut2);
			\vertex [above=0.2cm of cut2] (cut2u);
			\vertex [below=0.2cm of cut2] (cut2b);
			\diagram* {
			(b1)--[photon,ultra thick](u1), (b3)-- [photon,ultra thick] (u3b),(b3)-- [photon,ultra thick] (u3a), (p1) -- [thick] (b1)-- [thick] (b3)-- [thick] (p4), (cut1u)--[ red,thick] (cut1b),
			};
		\end{feynman}
  \end{tikzpicture}\\
  &+\begin{tikzpicture}[baseline={([yshift=-0.4ex]current bounding box.center)}]\tikzstyle{every node}=[font=\small]
		\begin{feynman}
			\vertex (p1) {\(\sc 5\)};
			\vertex [right=0.8cm of p1] (b1) [dot]{};
			\vertex [right=0.5cm of b1] (b2) []{};
			\vertex [right=0.5cm of b2] (b3) [HV]{H};
   \vertex [above=1.2cm of b1] (u1) []{$2$}; \vertex [above=1.2cm of b3] (u3) []{};\vertex [left=0.5cm of u3] (u3a) []{$1$};\vertex [right=0.5cm of u3] (u3b) []{$3$};
			\vertex [right=0.8cm of b3](p4){$\sc 4$};
			\vertex [right=0.5cm of b1] (cut1);
			\vertex [above=0.2cm of cut1] (cut1u);
			\vertex [below=0.2cm of cut1] (cut1b);
			\vertex [right=0.25cm of b2] (cut2);
			\vertex [above=0.2cm of cut2] (cut2u);
			\vertex [below=0.2cm of cut2] (cut2b);
			\diagram* {
			(b1)--[photon,ultra thick](u1), (b3)-- [photon,ultra thick] (u3b),(b3)-- [photon,ultra thick] (u3a), (p1) -- [thick] (b1)-- [thick] (b3)-- [thick] (p4), (cut1u)--[ red,thick] (cut1b),
			};
		\end{feynman}
  \end{tikzpicture}+\begin{tikzpicture}[baseline={([yshift=-0.4ex]current bounding box.center)}]\tikzstyle{every node}=[font=\small]
		\begin{feynman}
			\vertex (p1) {\(\sc 5\)};
			\vertex [right=0.8cm of p1] (b1) [dot]{};
			\vertex [right=0.5cm of b1] (b2) []{};
			\vertex [right=0.5cm of b2] (b3) [HV]{H};
   \vertex [above=1.2cm of b1] (u1) []{$3$}; \vertex [above=1.2cm of b3] (u3) []{};\vertex [left=0.5cm of u3] (u3a) []{$1$};\vertex [right=0.5cm of u3] (u3b) []{$2$};
			\vertex [right=0.8cm of b3](p4){$\sc 4$};
			\vertex [right=0.5cm of b1] (cut1);
			\vertex [above=0.2cm of cut1] (cut1u);
			\vertex [below=0.2cm of cut1] (cut1b);
			\vertex [right=0.25cm of b2] (cut2);
			\vertex [above=0.2cm of cut2] (cut2u);
			\vertex [below=0.2cm of cut2] (cut2b);
			\diagram* {
			(b1)--[photon,ultra thick](u1), (b3)-- [photon,ultra thick] (u3b),(b3)-- [photon,ultra thick] (u3a), (p1) -- [thick] (b1)-- [thick] (b3)-- [thick] (p4), (cut1u)--[ red,thick] (cut1b),
			};
		\end{feynman}
  \end{tikzpicture}+\begin{tikzpicture}[baseline={([yshift=-0.8ex]current bounding box.center)}]\tikzstyle{every node}=[font=\small]	
\begin{feynman}
    	 \vertex (a) {\( \bar 5\)};
    	 \vertex [right=1.2cm of a] (f2)[HV]{H};
    	 \vertex [right=1.2cm of f2] (c){$\bar 4$};
    	 \vertex [above=1.3cm of f2] (gm){$2$};
    	 \vertex [left=0.8cm of gm] (g2){$1$};
    	  \vertex [right=0.8cm of gm] (g20){$3$};
    	  \diagram* {
(a) -- [thick] (f2) --  [thick] (c),
    	  (g2)--[photon,ultra thick](f2),(gm)--[photon,ultra thick](f2),(g20)--[photon,ultra thick](f2)
    	  };
    \end{feynman}  
    \end{tikzpicture}\nn
\end{align}

\section{Spinning Compton Amplitudes}\label{section 3}

\subsection{Four-point Amplitude}\label{section 3.1}
We will now consider how to compute the classical four-point infinity-mass amplitude and start by writing the amplitude in terms of color-kinematic massless pole numerator factors (we define the shorthand notation  $p_{12}\equiv p_1 +p_2$):
\begin{align}\label{eq:Amp4}
A_a(12,v)&={\npre_a(1,2,v)-\npre_a(2,1,v)\over p_{12}^2}\equiv{\npre_a([1,2],v)\over p_{12}^2}\,.
\end{align}
Here $\npre_a(1,2,v)$ and $\npre_a(2,1,v)$ are numerators, and in the last step, we have used a commutator form $\npre_a([1,2],v)$ to write the amplitude in shorthand form.
The color-kinematic numerator $\npre_a([1,2],v)$ has a massive pole in $v\cdot p_1$ \cite{Brandhuber:2022enp}, and thus it follows from its factorization behavior that we have
\begin{align}
	\npre_a([1,2],v)\,\,\xrightarrow{} \,\, {p_1\cdot p_2\over m\,v\cdot p_1}\npre_a(1,v)\npre_a(2,v) = {1\over m\,v\cdot p_1}\,(p_1\cdot p_2) \,(w_1\cdot \veps_1) \, (w_2\cdot \veps_2)\,,
\end{align}
which we will rewrite as
\begin{align}\label{4-point}
	\npre_a([1,2],v)&=-{w_1\cdot F_1\cdot F_2\cdot w_2 \,\over m\,v\cdot p_1}+\npre'_a([1,2],v),
\end{align}
where $F_i^{\mu\nu} \equiv p_i^{\mu}\veps^{\nu}_i-\veps^{\mu}_ip^{\nu}_i$ denotes the abelian field strength. \\[5pt]
This separates the color-kinematic numerator into a $\displaystyle -{w_1\cdot F_1\cdot F_2\cdot w_2 \,\over m\,v\cdot p_1}$ with a pole, and an analytic and finite part $\npre'_a([1,2],v)$.\\[5pt]
Now factorizing the numerator for $p_{12}^2\to 0$ we see that 
\begin{align}\label{eq:fac4gluon}
    \npre_{a}([1,2],v)\big|_{p_{12}^2 \to 0}&= \sum_{\lambda_i}\npre_{\rm YM}([1,2],i)\times\npre_{a}(i,v),
\end{align}
where we can write the three-point Yang-Mills numerator as, see {\it e.g.} \cite{Brandhuber:2021bsf},
\begin{align}
	 \npre_{\rm YM}([1,2],i) &=-{\varepsilon_i\cdot F_1\cdot F_2\cdot \varepsilon_i\over \varepsilon_i\cdot p_1}. 
\end{align}
We consider an ansatz form to deduce $\npre'_a([1,2],v)$ at four points.
We first introduce the function $G_2(x_{1}; x_{2})$ (a generalization of the function $G_1(x_{1})$) where one again can prove that all apparent singularities are removable\footnote{\footnotesize{It is observed that the function $G_2(x_{2}; x_{1})$ is not independent since one can verify that 
$
	G_2(x_{2}; x_{1})+G_2(x_{1}; x_{2})=0.$  We stress that we are naturally led to the function $G_2(x_{1}; x_{2})$ when generating an amplitude ansatz free of spurious poles that satisfies the factorization requirements. Choosing an alternative ansatz function basis is possible but can be shown to affect only contact contributions not fixed by the considered factorizations.} }
\!\!\!\!\begin{align}G_2(x_{1}; x_{2})\equiv {1\over x_{2}}\Big({\sinh(x_{{12}})\over{x_{12}}}-\cosh(x_{2})\, {\sinh(x_{1})\over{x_1}}\Big)\equiv {1\over x_{2}}\Big(G_1(x_{12})-\cosh(x_{2}) G_1(x_1)\Big)\,.
\end{align} 
We then work out which types of terms we can have in the ansatz considering the above functions and evaluate the factorization in \Eq{eq:fac4gluon}. We find that any ansatz piece has to be of the following universal form (depending on the power in the spin variable),
\begin{align}
	\text{even}:&~ (v\cdot X\cdot a)(p\cdot Y\cdot a) \,G_1(x_{1})\,G_1(x_{2})\, \nn\\
	\text{odd}:&~ \tr(S\cdot X)\,G_1(x_{12}), \, (p\cdot X \cdot S\cdot Y\cdot a) \,G_2(x_{1};x_{2}),
\end{align}
where we have defined $x_i\equiv p_i\cdot a$ and $x_{i\ldots j}\equiv p_{i\ldots j}\cdot a$.
We follow the logic that we can only have a product of at most two $G$-functions since we only have two gluons at four points, and we insist on preserving the scaling in the spin variable of the three-point amplitudes by requiring that factors of $a$ present in the denominators of the $G$ functions cancel out with field strength monomial terms in the ansatz terms. Using the property of symmetry under parity ($a\rightarrow -a$) of the $G$-functions, it is furthermore easy to infer that only products $G_1(x_1)G_1(x_2)$ are possible for the even part. For the odd part, one can have either $G_1(x_{12})$ or $G_2(x_1;x_2)$ with a factor of $S$ or $...\cdot S\cdot X\cdot a$, respectively, with $X$ denoting a field strength tensor product. \\[5pt]
Thus, we consider all the 14 possible independent monomials in the ansatz and bootstrap its form using the fact that all massless factorizations are consistent. \newpage Thereby we entirely fix the factorization behavior in \Eq{eq:fac4gluon}, which leads to the expression:
\begin{align}\label{4-pointEntire}
	\npre'_a(12,v)&=m\,\Big((a\cdot F_2\cdot F_1\cdot v)\,( a\cdot p_1){-}(a\cdot F_2\cdot v )\,( a\cdot F_1\cdot p_2)\Big)\, G_1(x_1)\,G_1(x_2) \\
	&+\, i\,\Big(\big({a\cdot F_1\cdot F_2\cdot S\cdot p_2 } +a\cdot F_2\cdot F_1\cdot S\cdot p_1\big)\, G_2(x_1; x_2)\nn
 +\tr(S\cdot F_2\cdot F_1)G_1(x_{12})\Big).
\end{align}
To fix the factorization, we sum over gluon states using 
$$ \displaystyle  \sum_i \veps_i^{*\mu} \veps_i^{\nu}=\eta^{\mu\nu}.
$$
Although the form \Eq{4-pointEntire} is it is not manifestly anti-symmetric, one can straightforwardly check its anti-symmetry using the relation: 
\begin{align}
    &(a\cdot F_1\cdot p_2) \,(a\cdot F_2\cdot v)+(a\cdot F_2\cdot p_1) \,(a\cdot F_1\cdot v)\nn \\
&    -(a\cdot p_2) \,(a\cdot F_1\cdot F_2\cdot v)-(a\cdot p_1) \,(a\cdot F_2\cdot F_1\cdot v)=0.
\end{align}
Using the double copy relations, it follows that we have the classical spinning gravitational Compton amplitude:
\begin{align}\label{gravity4pt}
	\begin{tikzpicture}[baseline={([yshift=-15pt]current bounding box.center)}]\tikzstyle{every node}=[font=\small]	
\begin{feynman}
    	 \vertex (a) {\( \bar 4\)};
    	 \vertex [right=1.2cm of a] (f2)[HV]{H};
    	 \vertex [right=1.2cm of f2] (c){$\bar 3$};
    	 \vertex [above=1.3cm of f2] (gm){};
    	 \vertex [left=0.8cm of gm] (g2){$1$};
    	  \vertex [right=0.8cm of gm] (g20){$2$};
    	  \diagram* {
(a) -- [fermion,thick] (f2) --  [fermion,thick] (c),
    	  (g2)--[photon,ultra thick](f2),(g20)--[photon,ultra thick](f2)
    	  };
    \end{feynman}  
    \end{tikzpicture}\equiv M_a(12,v)={\npre_a([1,2],v)\,\npre_0([1,2],v)\over p_{12}^2},
\end{align}
where:
\begin{align}\label{4pointscalar}
\npre_0([1,2],v)&= -m\,\frac{v\cdot F_1\cdot F_2\cdot v }{v\cdot p_1}.
\end{align}\\[3pt]
As mentioned in the introduction, the expression for the Compton amplitude in gravity matches the recent literature \cite{Aoude:2022trd,Bautista:2022wjf,Bern:2020buy} up to quadratic order in spin. 

\subsection{Five-point Compton scattering}\label{Section 3.2}
We will now consider the five-point Compton amplitude. We can write this in the form
\begin{align}\label{eq:Amp5}
A_a(123,v)&={\npre_a([[1,2],3],v)\over p_{12}^2\,p_{123}^2}+{\npre_a([1,[2,3]],v)\over p_{23}^2\,p_{123}^2},
\end{align}
again expressed in terms of color-kinematic numerators solely organized in terms of the massless poles. Compared to the four-point case, we now have three massive poles in the color-kinematic numerators and thus arrive at the following factorization behavior: \\[3pt]
\begin{align}
	\npre_a([[1,2],3],v)&\rightarrow {p_{12}\cdot p_3\over m\,p_{12}\cdot v}\,\npre_a([1,2],v) \,\npre_a(3,v),\nn\\
	\npre_a([[1,2],3],v)&\rightarrow {p_{1}\cdot p_2\over m\,p_{13}\cdot v}\,\npre_a([1,3],v)\, \npre_a(2,v),\nn\\
	\npre_a([[1,2],3],v)&\rightarrow {p_{1}\cdot p_2\over m\,p_{1}\cdot v}\,\npre_a(1,v) \,\npre_a([2,3],v).\\[-10pt]\nn
\end{align}
Again as we did for the four-point case, we use factorization to recast the expression into one where we can identify the finite piece $\npre'_a([[1,2],3],v)$ with no pole\footnote{\footnotesize{For the part of the numerator with poles, we see that the first three terms follow a similar structure as in the spinless case \cite{Brandhuber:2021bsf} but with the change $v\cdot F_i \to w_i \cdot F_i$, which accounts for the inclusion of spin.}}
\begin{align}
&\npre_a([[1,2],3],v)=-{(w_1\cdot F_1\cdot F_2\cdot w_2) \, (p_{12}\cdot F_3\cdot w_3)\over m^2\,v\cdot p_1 \, v\cdot p_{12}}-{(w_1\cdot F_1\cdot F_3\cdot w_3) \, (p_{1}\cdot F_2\cdot w_2)\over m^2\, v\cdot p_1 \, v\cdot p_{13}}\nn\\
 &+{(w_1\cdot F_1\cdot F_2\cdot F_3\cdot w_3) \, \cosh(p_2\cdot a)\over m\,v\cdot p_1 }+{\npre'_a([1,2],v) \, (p_{12}\cdot F_3\cdot w_3)\over m\,v\cdot p_{12}}\nn\\
&+{\npre'_a([1,3],v) \, (p_{1}\cdot F_2\cdot w_2)\over m\,v\cdot p_{13}}+{(w_{1}\cdot F_1\cdot p_{2})\, \npre'_a([2,3],v) \, \over m\,v\cdot p_{1}}+\npre'_a([[1,2],3],v).
\end{align}
In this case, we find that it can be composed of the following building blocks (extending the logic from the four-point case) 
\begin{align}
	{\rm even}:\,& (v\cdot X \cdot a) \,(p\cdot Y\cdot a)\, G_1(x_i) \,G_1(x_{jk})\nn\\
 &(v\cdot X \cdot a)\, (p\cdot Y\cdot a)\,(p\cdot Z\cdot a)\, G_1(x_i)\, G_2(x_{j};x_k)\nn\\
{\rm odd}:\,& \tr(S\cdot X)\,  G_1(x_{123}),(p\cdot X\cdot S\cdot Y\cdot a)  \,G_2(x_{ij};x_k)\nn \\
 & (p\cdot X\cdot S\cdot Y\cdot a)\,(p\cdot Z\cdot a) \,G_3(x_{i};x_j,x_k) \nn \\
 & (p\cdot X\cdot S\cdot Y\cdot a)\,(p\cdot Z\cdot a) \,G_1(x_1)\,G_1(x_2)\,G_1(x_3),
\end{align}
where $\{i,j,k\}$ are elements in the possible permutations of the set $\{1,2,3\}$ and  
\begin{align}
G_3(x_{1}; x_{2},x_{3}):= {1\over x_{2} x_{3}}\Big({\sinh(x_{123})\over x_{123}}{-}\cosh(x_{2}) {\sinh(x_{13})\over x_{13}}\nn
 &{-}\cosh(x_{3}) 
  {\sinh(x_{{12}})\over x_{12}}\\{+}\cosh(x_{2}) \cosh(x_{3}){\sinh(x_{1})\over x_1}\Big)\,.
\end{align}
 As we discussed in the four-point case, we have some redundancy since the following type of identities hold 
\begin{align}
G_3(x_1;x_2,x_3)+G_3(x_2;x_1,x_3)+G_3(x_3;x_1,x_2)=-G_1(x_{1})\,G_1(x_{2})\,G_1(x_{3}).
\end{align}
The ansatz for the finite piece at five points has 981 independent (non-redundant) terms. Again we fix it completely by imposing the massless cut conditions,
\begin{align}
    A_{a}(123,v)\big|_{p_{123}^2 \to 0}&= \sum_{\lambda_i}A_{\rm YM}(123,i)\times\npre_{a}(i,v), \\
    \npre_{a}([[1,2],3],v)\big|_{p_{12}^2 \to 0}&= \sum_{\lambda_i}\npre_{\rm YM}([1,2],i)\times\npre_{a}([i,3],v),\nn\\
    \npre_{a}([[1,2],3],v)\big|_{p_{23}^2 \to 0}&= \sum_{\lambda_i}\npre_{\rm YM}([2,3],i)\times\npre_{a}([1,i],v),\nn
\end{align}
which leads to the following unique solution: 
\begin{equation}
    \npre'_a([[1,2],3],v)=\npre'_{\rm even}([[1,2],3],v)+\npre'_{\rm odd}([[1,2],3],v),
\end{equation}
where the spin-even part is 
\begin{align}
&\npre'_{\rm even}([[1,2],3],v){\,=\,}m\,\Big({-}\big((a\cdot F_1\cdot p_3) \, (a\cdot F_2\cdot p_1)\, (a\cdot F_3\cdot v)\big)\,G_1(x_3)\,G_2(x_{1}{;}x_{2})\nn\\
&{+}\big((a\cdot F_{12}\cdot p_3)\, (a\cdot F_3\cdot v){-}(p_{12}\cdot a)( a\cdot F_{312}\cdot v)\big)\,G_1(x_3)\,G_1(x_{12})\nn\\
&{-}\big((p_1\cdot p_2)\,(a\cdot F_1\cdot v)\,(a\cdot F_{32}\cdot a){+}(p_{23}\cdot a) (a\cdot F_1\cdot p_2)( a\cdot F_{32}\cdot v)\big) G_1(x_1)\,G_2(x_{2}{;}x_{3})\nn\\
&{+}(a\cdot F_2\cdot p_1) (a\cdot F_3\cdot F_1\cdot v) G_1(x_1)\,G_1(x_{2})\,\cosh(x_3)-(1{\leftrightarrow} 2)\Big)\nn\\
&{+}m\,((a\cdot F_1\cdot p_2) (a\cdot F_{23}\cdot v){-}(p_{1}\cdot a) (a\cdot F_{213}\cdot v)) \,G_1(x_1)\,G_1(x_2)\,\cosh(x_3)\,,
\end{align}
and the odd spin part is 
\begin{align}
    &\npre'_{\rm odd}([[1,2],3],v){\,=\,}\, i\,\Big(\tr(S\cdot F_{321})G_1(x_{123}){+} G_3(x_{1}{;}x_{2}{,}x_{3})\nn\\
    &
    \times\big((a\cdot F_2\cdot p_1) \big((a\cdot F_{31}\cdot S\cdot p_1){+}(a\cdot F_{13}\cdot S\cdot p_3)\big) {-} (p_1\cdot F_2\cdot S\cdot p_2)\, (a\cdot F_{31}\cdot a)\big)\nn\\&
   {-} (a\cdot F_{32}\cdot [S,F_1]\cdot p_{12})\,G_2(x_{12}{;}x_{3}){+}(a\cdot F_{21}\cdot [S{,} F_3]\cdot p_1)\, G_2(x_{13}{;}x_{2}){-}(1{\leftrightarrow} 2) \Big)\nn\\
     &+\, i\,\Big({-}(p_{1}\cdot a) (a\cdot F_{213}\cdot S\cdot p_3)\, G_3(x_{1}{;}x_{2}{,}x_{3}){-} (a\cdot F_{123}\cdot S\cdot p_3)\, G_2(x_{12}{;}x_{3})\nn\\
    &{-} (a\cdot F_{213}\cdot S\cdot p_3)\, G_2(x_{13};x_{2}){+} (a\cdot F_1\cdot p_2) \,(a\cdot F_{23}\cdot S\cdot p_3)\, G_1(x_{1})\,G_1(x_{2})\,G_1(x_{3}){\Big)}\,,
\end{align}
and we define $F^{\mu\nu}_{ij}\equiv F_i^{\mu\sigma} F_{j\sigma}^{\,\,\,\,\nu}$.
The appearance of $\cosh(x_i)$ arises from the linear combination of the $G$-functions depicted in \Eq{eq:recursive}. We have checked that the color-kinematic numerator satisfies the following manifest (see refs. \cite{Chen:2019ywi,Chen:2021chy}) crossing symmetry condition:
\begin{align}
   (\mathbb{I}{-}\mathbb{P}_{(21)})(\mathbb{I}{-}\mathbb{P}_{(321)})\,\npre_a([[1,2],3],v)=3 \,\npre_a([[1,2],3],v),
\end{align}
where the operator $\mathbb{I}$ is an identity operation and $\mathbb{P}_{(\tau)}$ denotes a cyclic permutation of gluon labels, {\it i.e.,} $\mathbb{P}_{(321)} f(1,2,3)\equiv f(3,1,2)$.\\[5pt]
Finally, we note that the classical spinning gravitational Compton amplitude can be obtained from the double copy:
\begin{align}
	&\begin{tikzpicture}[baseline={([yshift=-15pt]current bounding box.center)}]\tikzstyle{every node}=[font=\small]	
\begin{feynman}
    	 \vertex (a) {\( \bar 5\)};
    	 \vertex [right=1.2cm of a] (f2)[HV]{H};
    	 \vertex [right=1.2cm of f2] (c){$\bar 4$};
    	 \vertex [above=1.3cm of f2] (gm){$2$};
    	 \vertex [left=0.8cm of gm] (g2){$1$};
    	  \vertex [right=0.8cm of gm] (g20){$3$};
    	  \diagram* {
(a) -- [thick] (f2) --  [thick] (c),
    	  (g2)--[photon,ultra thick](f2),(gm)--[photon,ultra thick](f2),(g20)--[photon,ultra thick](f2)
    	  };
    \end{feynman}  
    \end{tikzpicture}\nn\equiv M_a(123,v)={\npre_a([1,[2,3]],v)\,\npre_0([1,[2,3]],v)\over p_{23}^2\,p_{123}^2}\\[5pt]
&\quad\quad+{\npre_a([[1,2],3],v)\,\npre_0([[1,2],3],v)\over p_{12}^2\,p_{123}^2}+{\npre_a([[1,3],2],v)\,\npre_0([[1,3],2],v)\over p_{13}^2\,p_{123}^2},
\end{align}
where the five-point numerator for the scalar case is \cite{Brandhuber:2021bsf}:
\begin{align}
        \mathcal{N}_0([[1,2],3],v) &= m\,\Big(\frac{v\cdot F_1\cdot F_2\cdot F_3 \cdot v}{v\cdot p_1}-\frac{v\cdot F_1\cdot F_2\cdot v \; p_{12}\cdot F_3\cdot v}{v\cdot p_1 v\cdot p_{12}} \nn\\
        &~~~~~-\frac{v\cdot F_1\cdot F_3\cdot v \; p_{1}\cdot F_2\cdot v}{v\cdot p_1 v\cdot p_{13}}\Big).
\end{align}%
\section{Building blocks for an all-multiplicity amplitude ansatz \black}\label{section 4}
Considering the four and five-point numerators, we will now make some general remarks about extensions of the formalism to all-multiplicity Compton amplitudes for massive classical spinning particles. Starting with amplitude representations with numerators organized in massless poles similar to the lower point examples, we can again decompose the color-kinematic numerators into pieces with massive poles and a finite term. We will focus mainly on bootstrapping the finite contribution. Similar to the lower point demonstrations, one can set up a convenient ansatz based on a generalization of the $G_1$, $G_2$, and $G_3$ functions. One can prove in all generality that the function $G_r(x_{1};x_{2},\cdots,x_{r})$ only features  removable singularities, see Appendix B.
\begin{align}
	G_r(x_{1};x_{2},{\cdots},x_{r})&\equiv {1\over x_{2}\cdots x_{r}}\label{generalGx}
 \times \Big(\sum\limits_{\tau_a\cup \tau_b=\tau}\sinch{x_{1\tau_a}}{\prod\limits_{i\in \tau_b}}({-}\cosh(x_i))\Big),
\end{align}
Here $\tau$ is the $\{2, \cdots, r\}$ and we have a summation over all the partitions of $\tau$ into subsets $\tau_a$ and $\tau_b$, including the empty set. It is easy to show that the following generic recursive relation holds
 \begin{align}\label{eq:recursive}
 G_r(x_{1};x_{2},\cdots,x_{r})={1\over x_r}\Big(G_{r-1}(x_{1}{+}x_r;x_{2},\cdots,x_{r-1})
 -G_{r-1}(x_{1};x_{2},\cdots,x_{r-1})\cosh(x_r)\Big).
 \end{align}
These functions also obey the following surprising relations:
\begin{equation}\label{sumG}
    \sum_{i=1}^n  G_n(x_i;x_1,\ldots,x_{i-1},x_{i+1},\ldots,x_n) = \begin{cases}
        0 & n\ \text{even}\\
        (-1)^{\frac{n-1}{2}}G_1(x_1)\ldots G_1(x_n)& n\ \text{odd}.
    \end{cases}
\end{equation}
The next step is to constrain the form of the finite term for the numerator. Again as in the lower point cases, to preserve the scaling in the spin variable of the three-point amplitudes, we require that the factors of $a$ present in the denominators of the $G$ functions cancel out with the elements in the monomials. In particular, if we define the degree of an object, {\it e.g.} 
\begin{align}
	\deg(G_r)=r,\ \deg(\cosh(x_j))=0 , \ \deg(a)=-1\,,
\end{align}
by its inverse scale order, ({\it i.e.,} the number of factors of $1/\lambda$ when taking $a\to \lambda a$  and $\lambda \to i \infty$).
Thus, for a general number of points, the following general ansatz for contributions is conjectured:
\begin{align}\label{ansatzGeneral}
	\begin{cases}
	(v\cdot X \cdot a )\mathcal{P}^{(r-1)}\mathcal{M}^{(r)}(G_1,\cdots, G_r)  &\text{even},  \\
	\tr(S\cdot X)G_1,\ (p\cdot X\cdot S\cdot Y \cdot a )\mathcal{P}^{(r-2)}\mathcal{M}^{(r)}(G_1,\cdots, G_r) & \text{odd}. 
\end{cases}
\end{align}
where $\mathcal{M}^{(r)}$ denotes a product of functions $G_i(x)$ such that the total degree is $r$, and $\mathcal{P}^{(r)}$ is a product of monomials of the arbitrary scalar products $p\cdot Z \cdot a$ with total degree $-r$ ($X,Y,Z$ all denote products of field strength tensors). The building blocks \Eq{ansatzGeneral} are all the possible combinations that respect the scaling in $a$ and that include the correct factors of $v$. 
These terms can be further constrained by considering parity requirements, so that for the even(odd) part, the ansatz should be even(odd) under the transformation $a\to -a$. \\[5pt]
After constructing the ansatz, the free parameters are fixed by the massless cut conditions:
\begin{align}
    \begin{tikzpicture}[baseline={([yshift=-0.8ex]current bounding box.center)}]\tikzstyle{every node}=[font=\small]    
   \begin{feynman}
    \vertex (a)[dot]{};
     \vertex [above=1.0cm of a](b)[HV]{\tiny YM};
     \vertex [above=0.4cm of a](bc)[]{};
     \vertex [left=0.5cm of bc](bcl)[]{};
     \vertex [right=0.5cm of bc](bcr)[]{};
     \vertex [left=1.1cm of b](c);
     \vertex [left=0.22cm of b](c23);
     \vertex [above=0.13cm of c23](v23)[]{};
    \vertex [above=.4cm of c](j1){$~$};
    \vertex [right=.9cm of j1](j2){$~~~~\cdots$};
    \vertex [right=1.3cm of j2](j3){$~$};
    \vertex [left=1.1cm of a](p1)[]{$~$};
    \vertex [right=1.1cm of a](p2)[]{$~$};
   	 \diagram*{(a) -- [gluon, thick] (b),(b) -- [gluon,thick] (j1),(b)--[gluon,thick](j3),(bcl)--[dashed,red,thick](bcr),(p1) -- [fermion, thick] (a)-- [fermion, thick] (p2)};
    \end{feynman}  
  \end{tikzpicture} &&
   \begin{tikzpicture}[baseline={([yshift=-0.8ex]current bounding box.center)}]\tikzstyle{every node}=[font=\small]    
   \begin{feynman}
    \vertex (a)[HV]{H};
     \vertex [above left=1.2cm of a](bt)[]{};
     \vertex [right=0.3cm of bt](b)[HV]{\tiny YM};
     \vertex [above right=2.3cm of a](j4t)[]{$\cdots~~~~~~~~~$};
     \vertex [left=0.9cm of j4t](j4)[]{};
     \vertex [right=0.9cm of j4](j5)[]{};
     \vertex [above left=0.6cm of a](bc)[]{};
     \vertex [above right=0.6cm of bc](bcl)[]{};
     \vertex [below left=0.6cm of bc](bcr)[]{};
     \vertex [left=0.8cm of b](c);
     \vertex [left=0.22cm of b](c23);
     \vertex [above=0.13cm of c23](v23)[]{};
    \vertex [above=.6cm of c](j1){$~$};
    \vertex [right=.6cm of j1](j2){$~~~~\cdots$};
    \vertex [right=0.9cm of j2](j3){$~$};
    \vertex [left=1.5cm of a](p1)[]{$~$};
    \vertex [right=1.5cm of a](p2)[]{$~$};
   	 \diagram*{(a) -- [gluon, thick] (b),(a) -- [gluon,thick] (j4),(a) -- [gluon,thick] (j5),(b) -- [gluon,thick] (j1),(b)--[gluon,thick](j3),(bcl)--[dashed,red,thick](bcr),(p1) -- [fermion, thick] (a)-- [fermion, thick] (p2)};
    \end{feynman}  
  \end{tikzpicture} 
\end{align}
The four and five points results indicate that our bootstrap procedure leads to a universal arbitrary-spin infinite-mass amplitude determined solely from exponentiated three-point amplitudes and factorization constraints. We expect this to be the case at any multiplicity.

\section{Conclusion}\label{section 5}
This paper discusses a new systematic bootstrap procedure for the computation of the leading in mass classical Compton scattering of arbitrary spinning infinity mass particles and gluons/gravitons that  avoid spurious poles and agrees with linear and quadratic results published in~\cite{Bern:2020buy,Bautista:2022wjf}. We have demonstrated this procedure by deducing new explicit classical results for four and five-point processes and discussed all multiplicity generalizations. \\[5pt]
Our results for four- and five-point scattering are suitable for concrete computations beyond this paper's scope; in particular, it opens an avenue for efficient calculation of the gravitational attraction of spinning black holes. For instance, by considering the following type of diagrammatic contributions illustrated below:
\begin{align}\label{eq:BA2PM}
	\begin{tikzpicture}[baseline={([yshift=-0.4ex]current bounding box.center)}]\tikzstyle{every node}=[font=\small]
		\begin{feynman}
			\vertex (p1) {\(v_1,a_1\)};
			\vertex [above=1.2cm of p1](p2){$v_2,a_2$};
			\vertex [right=1.3cm of p2] (u1) [HV]{H};
			\vertex [right=1cm of u1] (p3){};
			\vertex [right=0.8cm of p1] (b1) [dot]{};
			\vertex [right=0.5cm of b1] (b2) []{};
			\vertex [right=0.5cm of b2] (b3) [dot]{};
			\vertex [right=0.5cm of b3](p4){};
			\vertex [above=0.6cm of p1] (cutLt);
      \vertex [right=0.6cm of cutLt] (cutL);
			\vertex [right=1.55cm of cutL] (cutR){};
			\vertex [right=0.5cm of b1] (cut1);
			\vertex [above=0.2cm of cut1] (cut1u);
			\vertex [below=0.2cm of cut1] (cut1b);
			\vertex [right=0.25cm of b2] (cut2);
			\vertex [above=0.2cm of cut2] (cut2u);
			\vertex [below=0.2cm of cut2] (cut2b);
			\diagram* {
			(p2) -- [thick] (u1) -- [thick] (p3),
			(b1)--[photon,ultra thick](u1), (b3)-- [photon,ultra thick] (u1), (p1) -- [thick] (b1)-- [thick] (b3)-- [thick] (p4), (cutL)--[dashed, red,thick] (cutR), (cut1u)--[ red,thick] (cut1b),
			};
		\end{feynman}
  \end{tikzpicture}&& 
  \begin{tikzpicture}[baseline={([yshift=-0.4ex]current bounding box.center)}]\tikzstyle{every node}=[font=\small]
		\begin{feynman}
			\vertex (p1) {\(v_1,a_1\)};
			\vertex [above=1.2cm of p1](p2){$v_2,a_2$};
			\vertex [right=1.3cm of p2] (u1) [HV]{H};
			\vertex [right=1cm of u1] (p3){};
			\vertex [right=1.3cm of p1] (b1) [dot]{};
			\vertex [right=0.5cm of b1] (b2) []{};
			\vertex [right=0.5cm of b2] (p4) []{};
			\vertex [above=0.6cm of p1] (cutLt);
            \vertex [right=0.6cm of cutLt] (cutL);
			\vertex [right=1.55cm of cutL] (cutR){$k$};
			\vertex [right=0.5cm of b1] (cut1);
			\vertex [above=0.2cm of cut1] (cut1u);
			\vertex [below=0.2cm of cut1] (cut1b);
			\vertex [right=0.25cm of b2] (cut2);
			\vertex [above=0.2cm of cut2] (cut2u);
			\vertex [below=0.2cm of cut2] (cut2b);
			\diagram* {
			(p2) -- [thick] (u1) -- [thick] (p3),
			(b1)--[photon,ultra thick](u1), (cutR)-- [photon,ultra thick] (u1), (p1) -- [thick] (b1)-- [thick] (p4), (cutL)--[dashed, red,thick] (cutR),
			};
		\end{feynman}
	\end{tikzpicture} .
 \end{align}
we can compute non-spinning observables at the tree and one-loop post-Minkowskian order, see {\it e.g.} \cite{Brandhuber:2021eyq,Brandhuber:2021bsf,Brandhuber:2022enp} for examples using heavy-mass effective field theory or \cite{Bjerrum-Bohr:2021wwt,Bjerrum-Bohr:2022ows} using an analogous diagrammatic formulation in terms of velocity cuts. \\[5pt]
 Preliminary work on generalizations of the formalism presented suggests matching results beyond quadratic order (including Kerr, with the introduction of appropriate contact deformations) is feasible. It is beyond the scope of this paper, and we plan to return to it in a forthcoming publication. \\[5pt]
Another question is working out a connection between the numerators explored with three legs fixed and the formalism with numerators with two legs fixed used discussed in \cite{Bjerrum-Bohr:2019nws,Bjerrum-Bohr:2020syg} using momentum kernel identities that link the two formalisms \cite{Bjerrum-Bohr:2010pnr}.
While we have only computed the leading in mass part of the Compton amplitude, it is interesting to extend the results of this paper to the total amplitude, considering that the spinning particle has finite mass. To handle such amplitudes, one has to go beyond the simple factorization procedure regarded in the infinite mass limit. The reason is for high spin fields, where the mass is finite, there are different non-trivial effects from the complete spin propagator, which is a consequence of the massless particles interacting with the spinning particle and thereby altering its angular momentum. We will discuss this question further in appendix A.\\[5pt]
It is also an intriguing question to produce a Lagrangian formalism \cite{
Sorokin:2004ie,Metsaev:2005ar,Fotopoulos:2008ka,Buchbinder:2011xw,Ochirov_2022,Cangemi:2022bew} and kinematic algebra \cite{Monteiro:2011pc,Cheung:2016prv,Chen:2019ywi,Ben-Shahar:2021zww,Chen:2021chy,Chen:2022nei,Brandhuber:2021kpo,Brandhuber:2021bsf,Fu:2016plh, Reiterer:2019dys,Tolotti:2013caa,Mizera:2019blq,Borsten:2020zgj,Borsten:2020xbt,Ferrero:2020vww,Borsten:2021hua,Cheung:2021zvb,Ben-Shahar:2022ixa} for numerators that potentially could supply a guideline for a general and fundamental framework for Compton scattering valid to all orders in spin.  We leave this question for future work.
\section*{Acknowledgements}
G.C. would like to thank Jung-Wook Kim for several profound discussions. We also thank Yilber F. Bautista, Marco Chiodaroli, Poul H. Damgaard, Henrik Johansson, Kays Haddad, Yutin Huang, Andres Luna and Alexander Ochirov for their comments.   The work of N.E.J.B.-B. and G.C. was supported by DFF grant 1026-00077B and in part by the Carlsberg Foundation. G.C. has received funding from the European Union Horizon 2020 research and innovation program under the Marie Sklodowska-Curie grant agreement No. 847523 INTERACTIONS. M.S. was supported in part by the Stefan and Hanna Kobylinski Rozental Foundation.


\appendix
\section{Classical spin theory with finite mass}
Extending the amplitudes for classically spinning particles beyond the infinite mass limit is highly non-trivial since all the possible spin states for the internal particles in the propagators have to be  summed over. Such effects are sub-leading when doing an expansion in the mass  $m$~\cite{Bautista:2021wfy,Bautista:2022wjf}, which is why we did not consider them previously.  However, for the sake of clarity, we will compute the piece of the full expression with a non-zero contribution in the Heavy-mass Effective Field Theory. \\[5pt]
We start at three points, where the numerator has been calculated \cite{Arkani-Hamed:2017jhn}  
\begin{align}
\floor{\npre_a(1,\bar 2,\bar 3)}=\floor{w^{(3)}_{1}}\cdot \veps_1\,.
\end{align}
Here $(\floor{w^{(j)}_{i}})^{\mu}= \floor{\cosh(x_i)} p_j^{\mu}-\, i\, \,\floor{\sinch{x_i}}(p_i\cdot  S_{j})^{\mu}$ and $S^{\mu\nu}_j=-\epsilon^{\mu\nu\rho\sigma} (p_{j})_{\rho} a_\sigma$. The $\floor{~}$ denotes formally truncated pieces in spin order. In the following, we will omit all the $\floor{~}$ symbols and work under the assumption that we have formally expanded to some finite order. As in the main text, we denote all massive particles with a bar.\\[5pt]
For the four-point amplitude, one needs to consider the propagator of the finite spin theory, as studied in \cite{Scadron:1968zz,Bautista:2022wjf}. However, if we are only interested in the pieces of the amplitude which contribute to the heavy mass limit, then we can separate it as
\begin{align}
	A^{({\rm total})}_a(1,2,\bar 3, \bar 4)=A_a(1,2,\bar 3, \bar 4)+ \text{sub-leading finite spin effect},
\end{align}
where the additional contribution is sub-dominant in the $1/m$ expansion.
\newpage Assuming that $A_a(1,2,\bar 3, \bar 4)$ satisfies the color-kinematics duality, we can write it in numerator form
\begin{align}
	A_a(1,2,\bar 3, \bar 4)={\npre_a(1,2,\bar 3,\bar 4)\over p_{12}^2}, 
\end{align}
and we will here construct the color-kinematic numerator directly. \\[5pt]
First, we note that from the factorization behavior in the massive pole, we have 
\begin{align}
\npre_a(1,2,\bar 3,\bar 4)=-{w^{(4)}_{1}\cdot F_1\cdot F_2\cdot w^{(4)}_{2}\over p_4\cdot p_1}+{p_4\cdot p_2 \over p_4\cdot p_1}\npre'_{1}(1,2, \bar 3,\bar 4)+\npre'_{2}(1,2, \bar 3,\bar 4).
\end{align}
It is then easy to see that the first term factorizes into a product of two 3-point amplitudes in the cut
\begin{align}
-{w^{(4)}_{1}\cdot F_1\cdot F_2\cdot w^{(4)}_{2}\over p_4\cdot p_1}\xrightarrow{p_4\cdot p_1 \to 0} {p_1\cdot p_2 \over p_4\cdot p_1}w^{(4)}_{1}\cdot\veps_1   w^{(4)}_{2}\cdot\veps_2.
\end{align}
The second term is a new non-trivial contribution from the finite spin effect. The factor $p_4\cdot p_2/p_4\cdot p_1$ is the only possible object that transforms into a non-singular function in the heavy mass limit, {\it i.e.}, $p_4\cdot p_2 \to v\cdot p_2 = -v\cdot p_1$. We see that the last term fixes the factorization behavior for the massless pole term. \\[5pt]
Similarly to what happens in the heavy mass limit, the denominator of the singular term cancels when taking the cut $p_1\cdot p_2 \to 0$
\begin{align}
-{w^{(4)}_{1}\cdot F_1\cdot F_2\cdot w^{(4)}_{2}\over p_4\cdot p_1}\Big|_{p_1\cdot p_2 \to 0}= -{\cosh(x_1)} \veps_1\cdot F_2 \cdot w^{(4)}_{2}+w^{(4)}_{1}\cdot\veps_1   p_1\cdot\veps_2 {\cosh(x_2)}.
\end{align}
It follows that from the factorization behavior on the massless pole, 
\begin{align}
\npre_a(1,2,\bar 3,\bar 4)|_{p_1\cdot p_2 \to 0}= -\veps_1\cdot F_2\cdot w^{(4)}_{12}+ w^{(4)}_{12}\cdot \veps_1 \, p_1\cdot \veps_2,
\end{align}
we have 
\begin{align}
  &  -\npre'_1(1,2,\bar 3,\bar 4)+\npre'_2(1,2,\bar 3,\bar 4)|_{p_1\cdot p_2 \to 0}\nn \\&=-\veps_1\cdot F_2\cdot w^{(4)}_{12}+ w^{(4)}_{12}\cdot \veps_1 \, p_1\cdot \veps_2-\Big(w^{(4)}_{1}\cdot\veps_1   p_1\cdot\veps_2 {\cosh(x_2)}-{\cosh(x_1)} \veps_1\cdot F_2 \cdot w^{(4)}_{2}\Big)\nn\\
    &=(a\cdot p_2 a\cdot p_1 \veps_1\cdot F_2\cdot p_4 -a\cdot p_2 a\cdot p_1 p_1\cdot \veps_2 \veps_1\cdot p_4)G_1(x_1)G_1(x_2)\nn\\
    &+\, i\, (- \veps_1\cdot F_2\cdot S_4\cdot p_1 +p_1\cdot \veps_2 \veps_1\cdot S_4\cdot p_2)G_1(x_{12})\nn\\
    &+\, i\,(a\cdot p_1 \veps_1\cdot F_2\cdot S_4\cdot p_2+p_1\cdot \veps_2 a\cdot p_2 \veps\cdot S_4\cdot p_1)G_2(x_1;x_2).
\end{align}
Now, one can show that to satisfy the color-kinematics duality (i.e., the Jacobi relations between the half-ladder numerators), the following must hold 
\begin{align}
\npre'_1(1,2,\bar 3,\bar 4)=- \npre'_2(1,2,\bar 3,\bar 4)\equiv {1\over 2}\npre'_a(1,2,\bar 3,\bar 4),
\end{align}
so that  
\begin{align}
	\npre'_a(1,2,\bar 3,\bar 4)&=\Big({p_4\cdot p_2-p_4\cdot p_1\over 2} \, a\cdot F_1\cdot F_2\cdot a +a\cdot F_1\cdot p_2 \, a\cdot F_2\cdot p_4\nn \\ & -a\cdot F_2\cdot p_1 \, a\cdot F_1\cdot p_4\Big)G_1(x_1)G_1(x_2)\nn\\
 &+\, i\, (a\cdot F_1\cdot F_2\cdot S_4\cdot p_2 +a\cdot F_2\cdot F_1\cdot S_4\cdot p_1 )G_2(x_1;x_2)\nn\\
 &+\, i\,\tr(F_1\cdot F_2\cdot S_4) G_1(x_{12})
\end{align}
is determined by requirements of gauge invariance and the factorization behavior of massless pole terms. We conclude by expressing the full numerators as follows
\begin{align}
\npre_a(1,2,\bar 3,\bar 4)=-{w^{(4)}_{1}\cdot F_1\cdot F_2\cdot w^{(4)}_{2}\over p_4\cdot p_1}+{p_4\cdot p_1-p_4\cdot p_2 \over 2p_4\cdot p_1}\npre'_a(1,2, \bar 3,\bar 4).
\end{align}
A direct check of the color-kinematic duality is from the BCJ relations with a massive particle \cite{Johansson:2015oia}
\begin{align}
	p_4\cdot p_1 A_a(1,2,\bar 3,\bar 4)=p_4\cdot p_2 A_a(2,1,\bar 3,\bar 4). 
\end{align}
Finally, we obtain the gravity amplitude by performing the double copy with a scalar Compton amplitude \cite{Bern:2020buy,Johansson:2019dnu} yielding
\begin{align}\label{eq:GRAmp4}
M_a(1,2,\bar 3,\bar 4)=-{1\over p_{12}^2}\npre_a(1,2,\bar 3,\bar 4)\npre_0(1,2,\bar 4,\bar 3),
\end{align}
where $\npre_0$ is the color-kinematic numerator for the interaction of a massive scalar line with two gluons \cite{Brandhuber:2021eyq,Chen:2022nei}:
\begin{align}
   \npre_0(1,2,\bar 4,\bar 3)= -{p_3\cdot F_1\cdot F_2\cdot p_3\over p_3\cdot p_1}.
\end{align}
We can check the consistency of the derived numerators and their corresponding infinite mass effective theory expressions. We see that
\begin{align}
\npre_0(1,2,\bar 4, \bar 3)|_{m\rightarrow \infty}&=-\npre_0([1,2],v)\nn\\
\npre_a(1,2,\bar 3, \bar 4)|_{m\rightarrow \infty}&=\npre_a([1,2],v)
\end{align}
where $\npre_a([1,2],v)$ is given in \Eq{4-point} and \Eq{4-pointEntire}, and $\npre_0([1,2],v)$ by \Eq{4pointscalar}.\\[5pt]
In order to see that this amplitude has the correct factorization behaviour on the massive cut, we first take the limit $p_4\cdot p_1 \to 0$ in \Eq{eq:GRAmp4}, and only then perform an expansion in $1/m$, obtaining the following result: 
\begin{align}
    M_a(1,2,\sc 3,\sc 4)&\xrightarrow {p_4\cdot p_1 \rightarrow 0}{\Big(p_4\cdot \veps_1 p_4\cdot \veps_2+\mathcal{O}({1\over m})\Big)\Big(w^{(4)}_{1}\cdot\veps_1   w^{(4)}_{2}\cdot\veps_2+\mathcal{O}({1\over m})\Big) \over p_4\cdot p_1}\nn\\
    &\xrightarrow{m\rightarrow \infty} m^3{M_a(1,v)M_a(2,v)\over v\cdot p_1}.
\end{align}
This is consistent with our reasoning in section \ref{AmpDecomp} since, taking into account the regulators $i\varepsilon$, the decomposition of the amplitude contains a term of order $\mathcal{O}(m^3)$ where the massive cut is imposed by a $\delta$-function. For example, the bottom graph in \Eq{eq:BA2PM} is deduced from
\begin{align}\label{eq:cut}
	\begin{tikzpicture}[baseline={([yshift=-0.4ex]current bounding box.center)}]\tikzstyle{every node}=[font=\small]
		\begin{feynman}
			\vertex (p1) {\(\sc 4\)};
			\vertex [right=0.8cm of p1] (b1) [dot]{};
			\vertex [right=0.5cm of b1] (b2) []{};
			\vertex [right=0.5cm of b2] (b3) [dot]{};
   \vertex [above=1.2cm of b1] (u1) []{$1$}; \vertex [above=1.2cm of b3] (u3) []{$2$};
			\vertex [right=0.8cm of b3](p4){$\sc 3$};
			\vertex [right=0.5cm of b1] (cut1);
			\vertex [above=0.2cm of cut1] (cut1u);
			\vertex [below=0.2cm of cut1] (cut1b);
			\vertex [right=0.25cm of b2] (cut2);
			\vertex [above=0.2cm of cut2] (cut2u);
			\vertex [below=0.2cm of cut2] (cut2b);
			\diagram* {
			(b1)--[photon,ultra thick](u1), (b3)-- [photon,ultra thick] (u3), (p1) -- [thick] (b1)-- [thick] (b3)-- [thick] (p4),
			};
		\end{feynman}
  \end{tikzpicture}+\begin{tikzpicture}[baseline={([yshift=-0.4ex]current bounding box.center)}]\tikzstyle{every node}=[font=\small]
		\begin{feynman}
			\vertex (p1) {\(\sc 4\)};
			\vertex [right=0.8cm of p1] (b1) [dot]{};
			\vertex [right=0.5cm of b1] (b2) []{};
			\vertex [right=0.5cm of b2] (b3) [dot]{};
   \vertex [above=1.2cm of b1] (u1) []{$1$}; \vertex [above=1.2cm of b3] (u3) []{$2$};
			\vertex [right=0.8cm of b3](p4){$\sc 3$};
			\vertex [right=0.5cm of b1] (cut1);
			\vertex [above=0.2cm of cut1] (cut1u);
			\vertex [below=0.2cm of cut1] (cut1b);
			\vertex [right=0.25cm of b2] (cut2);
			\vertex [above=0.2cm of cut2] (cut2u);
			\vertex [below=0.2cm of cut2] (cut2b);
			\diagram* {
			(b1)--[photon,ultra thick](u3), (b3)-- [photon,ultra thick] (u1), (p1) -- [thick] (b1)-- [thick] (b3)-- [thick] (p4), ,
			};
		\end{feynman}
  \end{tikzpicture}=\begin{tikzpicture}[baseline={([yshift=-0.4ex]current bounding box.center)}]\tikzstyle{every node}=[font=\small]
		\begin{feynman}
			\vertex (p1) {\(\sc 4\)};
			\vertex [right=0.8cm of p1] (b1) [dot]{};
			\vertex [right=0.5cm of b1] (b2) []{};
			\vertex [right=0.5cm of b2] (b3) [dot]{};
   \vertex [above=1.2cm of b1] (u1) []{$1$}; \vertex [above=1.2cm of b3] (u3) []{$2$};
			\vertex [right=0.8cm of b3](p4){$\sc 3$};
			\vertex [right=0.5cm of b1] (cut1);
			\vertex [above=0.2cm of cut1] (cut1u);
			\vertex [below=0.2cm of cut1] (cut1b);
			\vertex [right=0.25cm of b2] (cut2);
			\vertex [above=0.2cm of cut2] (cut2u);
			\vertex [below=0.2cm of cut2] (cut2b);
			\diagram* {
			(b1)--[photon,ultra thick](u1), (b3)-- [photon,ultra thick] (u3), (p1) -- [thick] (b1)-- [thick] (b3)-- [thick] (p4), (cut1u)--[ red,thick] (cut1b),
			};
		\end{feynman}
  \end{tikzpicture}.
  \end{align} 

\section{Removable singularities in the $G$ functions}
We will here present the proof that the $G_n(x_{1};x_{2},{\cdots},x_{n})$ in \Eq{generalGx} functions 
only feature removable singularities, {\it i.e.} when expanding these functions as power series around $x_1=...=x_n=0$,  all poles are removable, leading to a finite result at $x_1=...=x_n=0$. We will prove this by induction. For $n=2$, it is straightforward to check this property simply by performing a Taylor expansion and seeing that the singular term cancels out.\\[5pt]
Define now $\Tilde{G}_n(x_1;x_2,\ldots,x_n):=x_2\cdots x_nG_n(x_1;x_2,\ldots,x_n)$. Assuming that the statement holds for $n-1$, we have
\begin{align}
    \Tilde{G}_{n-1}(x_{1n};...,x_{n-1})&=\sinch{x_{1\ldots n-1}} - \sum_{i=2}^n\sinch{x_{1\ldots i-1}+x_{i+1\ldots n-1}}\cosh(x_i)\\ 
    &+ \sum_{i<j=2}^n\sinch{x_{1\ldots i-1}+x_{i+1\ldots j-1}+x_{j+1\ldots n-1}}\cosh(x_i)\cosh(x_j)\nn\\
    & +\cdots +(-1)^{n-2}\sinch{x_{1n}}\cosh(x_2)...\cosh(x_{n-1}) \propto x_2\ldots x_{n-1}.\nn
\end{align}
The remaining terms are
\begin{align}
    &\Tilde{G}_n(x_1;x_2,\ldots,x_n)- \Tilde{G}_{n-1}(x_{1n};x_2,\ldots,x_{n-1})=-\sinch{x_{1\ldots n-1}}\cosh(x_n)  \nn \\
    &+\sum_{i=2}^n\sinch{x_{1\ldots i-1}+x_{i+1\ldots n-1}}\cosh(x_i)\cosh(x_n)+ \cdots + (-1)^{n-1}\sinch{x_1}\prod_{i=2}^{n}\cosh(x_i)\nn \\
    &= -\Tilde{G}_{n-1}(x_1,\cdots,x_{n-1})\cosh(x_n) \propto x_2...x_{n-1}.
\end{align}
In other words, both contributions are proportional to $x_2\ldots x_{n-1}$. Now, if we group the terms in the previous two equations by pairs, we get the $n=2$ statement
\begin{align}
      &  \sinch{x_{1\ldots n}}-\sinch{x_{1\ldots n-1}}\cosh(x_n) =  \Tilde{G}_2(x_{1\ldots n-1};x_n)\propto x_n, \nn\\
      &  -\Big(\sinch{x_{1\ldots i-1}+x_{i+1\ldots n}}\cosh(x_i)- \sinch{x_{1\ldots i-1}+x_{i+1\ldots n-1}}\cosh(x_i)\cosh(x_n)\Big)\nn\\ 
       & = -\Tilde{G}_2(x_{1\ldots i-1}+x_{i+1\ldots n-1};x_n)\cosh(x_i)\propto x_n,\nn \\
       & ~~~~~~~~~~~~~~~~~~~~~~~~~~~~~~~\vdots~~~~~~~~~~~~~~~~~~~~~~~~~~~~~~~\vdots \\
       & \left( \sinch{x_{1n}}\prod_{i=2}^{n-1}\cosh(x_i)- \sinch{x_1}\prod_{i=2}^{n}\cosh(x_i) \right)
        = \Tilde{G}_{2}(x_1;x_n)\prod_{i=2}^{n-1}\cosh(x_i) \propto x_n.\nn
\end{align}
With this, we have proven that $\Tilde{G}_n(x_1,\ldots,x_n)$ is both proportional to $x_2\cdots x_{n-1}$ and to $x_n$ or, equivalently, that it is proportional to $x_2\cdots x_n$. 

\bibliographystyle{JHEP}
\bibliography{KinematicAlgebra}
\end{document}